\documentclass[runningheads]{llncs}

\usepackage{subcaption}
\usepackage{algorithm}
\usepackage{algpseudocode}
\usepackage{graphicx}
\usepackage{amsmath}

\usepackage{comment}
\includecomment{doubleblind}
\excludecomment{doubleblindreplacement}

\usepackage[final]{changes}

\begin{doubleblind}

\end{doubleblind}

\begin{doubleblindreplacement}

\end{doubleblindreplacement}

\newcommand{\teaMPI}{teaMPI}

\usepackage{url}

\usepackage[absolute]{textpos}

\begin{document}
  \title{TeaMPI---Replication-based Resilience without the (Performance) Pain}

\begin{doubleblind}

\author{
  Philipp Samfass\inst{1}
  \and
  Tobias Weinzierl\inst{2}
  \and
  Benjamin Hazelwood\inst{2}
  \and
  Michael Bader \inst{1}
}

\authorrunning{P. Samfass et al.}

\institute{
 Technische Universit\"at M\"unchen,
 85748 Garching, Germany
 \email{\{samfass,bader\}@in.tum.de}
 \and
 Computer Science,
 Institute for Data Science,
 Durham University,
 DH13LE Durham, Great Britain\\
 \email{tobias.weinzierl@durham.ac.uk, ben.hazelwood@featurespace.co.uk}
}
\end{doubleblind}

\begin{doubleblindreplacement}

\author{
  Author1\inst{1}
  \and
  Author2\inst{2}
  \and
  Author3\inst{3}
  \and
  Author4\inst{4}
}

\authorrunning{Author1 et al.}

\institute{
 Omitted due to double blind policy,
 Omitted due to double blind policy
 \email{anonymous@email.com}
 \and
 Omitted due to double blind policy,
 Omitted due to double blind policy
 \email{anonymous@email.com}
 \and
 Omitted due to double blind policy,
 Omitted due to double blind policy
 \email{anonymous@email.com}
}

\end{doubleblindreplacement}

  \maketitle

  \begin{abstract}
    In an era where we can not afford to checkpoint frequently, replication
is a generic way forward to construct numerical
simulations that can continue to run even if hardware parts fail.
Yet, replication often is not employed on larger scales, as na\"ively
mirroring a computation once effectively halves the machine size,
and as keeping replicated simulations consistent with each other is
not trivial.
We demonstrate for the \added{ExaHyPE engine --- a task-based solver for hyperbolic
equation systems ---} that it is possible to realise resiliency without
major code changes on the user side, while we 
introduce a novel algorithmic idea where replication reduces the
time-to-solution.
The redundant CPU cycles are not burned ``for nothing''.
Our work employs a weakly consistent data model where replicas run 
independently yet inform each other through heartbeat messages whether they are still up and
running.
Our key performance idea is to let the tasks of the
replicated simulations share some of their outcomes, while we shuffle the actual
task execution order per replica.
This way, replicated ranks can skip some  
local computations and automatically start to synchronise with each
other.
Our experiments with a production-level \added{seismic} wave-equation solver provide evidence
that this novel concept has the potential to make replication affordable for
large-scale simulations in high-performance computing.

  \end{abstract}


\begin{textblock}{12}(2.0,0.93) 
\noindent\footnotesize
This paper has been published in the proceedings of the ISC 2020.\\
The final authenticated version is available online on \url{https://doi.org/10.1007/978-3-030-50743-5_23}.
\end{textblock}

\section{Introduction}\label{section:introduction}

%
%
Supercomputing roadmaps predict that machines soon will suffer from 
hardware unreliability \cite{Dongarra:14:ApplMathExascaleComputing}. 
A linear correlation between system 
size and the number of failures has already been observed
\cite{Schroeder:2010:LargeScale}, as effects alike bias
temperature instabilities or hot carrier injection diminish the mean time between failures
(MTBF) for the individual components.
For the next generation of machine sizes,
a preserved or reduced 
MTBF however implies that codes have to be prepared for parts of the
machine going down unexpectedly, either through hard errors or soft errors
corrupting the code's state.
Alternatively, parts might become unacceptably slow as hardware or software
error correction \cite{Reinarz:18:APosterioriFaultFrequency} step in\deleted{ to preserve a correct behaviour}.
We thus need resilient codes.
Numerical simulations will have to be at the forefront here.
With their massive concurrency going full speed and their strong causal
dependencies between intermediate results they are 
vulnerable to hardware failures.

%
%
For numerical simulations, we distinguish three \deleted{different }strategies to inject
resilience: 
(i)~Codes can \replaced{be prepared algorithmically}{algorithmically be prepared} to recover from drop-outs of compute
nodes.
(ii)~Codes can checkpoint and restart if hardware fails.
(iii)~Codes can run computations redundantly.

The first variant works only if the underlying problem allows us to
recover information even if data is ``lost''.
Elliptic equations fall into this category: If we know the solution around a region
that has dropped out, we can reconstruct the solution within the domain
\cite{Altenbernd:2018:FaultsInMultigrid,Goddeke:2015:FaultTolerantMultigrid}.
Another example for algorithmic recovery is the combination technique, where a
drop-out of some data might (slightly) reduce the solution accuracy but the
overall algorithm can cope with it \cite{Heene:17:SparseGridsFaultTolerance}.
In both cases, 
the numerical scheme itself has to be resiliency-ready.

Checkpointing works more in a black-box fashion, but the time to write a
checkpoint has to be significantly smaller than the MTBF.
We also have to be willing to spend CPU cycles and energy on I/O, which
typically is costly \cite{Ferreira2011a}.
For in-memory checkpointing which mitigates the speed and energy penalty, we
need ``spare'' storage.
As checkpoints are costly one way or the other, partial checkpoint-restart is a
must.
Containment domains \cite{chung12containment} for example ask the
programmer to decompose the application into task-similar constructs with
manual state preservation, error detection and recovery.
Some tasking runtime systems such as ParSEC \cite{cao15sdcruntime} provide
a framework for the ``automatic'' re-execution of task sub-graphs in combination
with checkpointing.
A sophisticated example for checkpointing is to run the recalculation with a
different numerical scheme \cite{Reinarz:18:APosterioriFaultFrequency}.
This realises a hybrid
between an algorithmic approach and checkpoints.
%

If algorithmic resiliency is not at hand and checkpointing cannot be afforded,
replication of work, i.e.,~data
redundancy, is the prime solution.
If a node or memory drops out, we simply swap in the replicated data.
Cloud computing, sensor networks, desktop grids, peer-to-peer
networks, and almost every other field that requires resilient
computations \cite{Cappello2009}
base their fault tolerance upon the idea of
replicating resources.
Capability high-performance computing (HPC) in contrast 
tends not to use replication.
If we duplicate a computation, we effectively half the machine---which
renders the prime character of capability computing absurd.
Since supercomputers however tend to become so ill-balanced
w.r.t.~I/O~capabilities vs.~compute resources that we cannot afford to
checkpoint frequently, we will eventually be forced to employ replication
nevertheless \cite{Engelmann2009,engelmann2014scaling,Riesen2010}.
\replaced{We therefore need}{As a result, a concerted effort is required} to reduce its pain.
\deleted{%
We introduce a novel idea to do so, together with a prototypical
implementation called \teaMPI. 
We demonstrate its potential for a high-order discontinuous Galerkin code 
for hyperbolic equation systems \cite{Software:ExaHyPE}, i.e.,~a solver for which
we are not aware of any straightforward algorithmic resiliency strategy.
}


%
%
%
\added{%
Our paper introduces a novel idea to do so, together with a prototypical
implementation of team-based MPI replication, called \teaMPI. 
We demonstrate its potential for a high-order discontinuous Galerkin code 
for hyperbolic equation systems, i.e.,~a solver for which we are not aware of any
straightforward algorithmic resiliency strategy.
}
Our \replaced{approach}{paper} relies on replication \deleted{operating }on the MPI rank
level.
Each rank is replicated $K$ times, while the
simulation per rank is phrased in tasks.
A task is an atomic unit, i.e.,~it has a well-defined input and output and, once
it becomes ready, can be executed without any further dependencies.
To benefit from our techniques, 
\replaced{%
codes need not be task-based only, 
}{%
it is not necessary for a code to run task-based only, 
}%
but the heavy workload should be phrased as tasks.
\added{Furthermore, we require that tasks allow us
to send their outcome via MPI, and they should have some slack,
i.e.,~should not be part of the critical path.
That is, there is some freedom to move their startup time around, without
immediately penalising the overall time-to-solution.
}

\replaced{%
With such tasks, we can replicate each rank $K$ times 
without a $K\times$ overhead in compute time:
}{%
With our new idea, the total compute-time cost
is not increased by a factor of $K$ even though we replicate each rank $K$ times:
}%
We shuffle the task execution order per replication,
i.e.,~we make each rank process sets of ready tasks in a slightly different
order. 
Furthermore, we \replaced{let}{give} each rank \deleted{the opportunity to }offer its task outcomes to
other replicas.
Whenever a task is about to be executed on a \deleted{particular }rank, this
rank now can first check whether the task outcome is already available from a replica.
If so, it skips the \deleted{task }execution.
All techniques \replaced{affect}{are realised within} the task runtime,
i.e.,~\replaced{can be}{are completely} invisible to the \deleted{actual} simulation
\cite{Subasi:2017:SelectiveReplication}.
To the best of our knowledge, this is the first approach offering full
simulation replication without a full multiplication of compute workload.
It is thus a fundamental contribution towards affordable replication-based
resiliency in HPC.

%
%
\replaced{%
While our paper focuses on speed of execution, we can detect certain hard failures as well, 
using a concept called \emph{heartbeats}.
}{%
Our paper provides a
feasibility study focusing on speed of execution.
Using a concept called heartbeats, we however can detect timeouts as well: 
}%
Since we keep redundant copies of data, we could, similar to RAID systems,
replace corrupt ranks.
However, a discussion and presentation of such a swapping strategy is beyond the scope of this paper.
Furthermore, we do not yet link our work to MPI-based
run-through-stabilisation techniques
\cite{bland2013ulfm,chen2005fault,fagg2005process}, 
which inject further technical and implementation difficulties. 
Finally, replication in HPC remains a double-edged sword: 
\replaced{While}{On the one hand,} it offers fault-tolerance\replaced{,}{.} 
\replaced{it also}{On the other hand, it} requires to use more memory, network
bandwidth and compute units, i.e.,~CPU hours, per simulation run. 
Our approach reduces the compute cost compared to na\"ive replication.
We however neglect the increased memory \cite{Biswas:11:DataSimilarity} and
network stress.
For many applications, users will have to balance the replication-based resilience
against these facets of increased cost.

The remainder of the text is organised as follows:
We establish our terminology in Section \ref{section:team-based-resiliency} and
sketch the replication mechanisms.
Our core contribution is the introduction of the task-based result sharing
(Section \ref{section:task-sharing}) which eventually reduces the
workload per rank whenever results from a replica drop in on time.
The realisation and encapsulation of the whole idea is subject of discussion in
Section \ref{section:implementation}, before we study the method's runtime
implications (Section \ref{section:results}).
A brief conclusion and outlook in Section \ref{section:conclusion} wrap up the
discussion.


\section{Team-based resiliency with heartbeats}
\label{section:team-based-resiliency}

We first introduce the terminology which underlies our
algorithmic contributions.
The terminology is also adopted by our software \teaMPI\ which realises the
proposed ideas.
\teaMPI\ plugs into the MPI standard profiling interface (PMPI).
By mapping physical ranks onto logical numbers and altering the MPI communicator size
(number of ranks) exposed to SPMD user code, it transparently reorganises and 
replicates MPI
ranks in multiple teams:

%
%
\begin{definition}[Team]
All the ranks of an application (without any redundancy) form a team.
If we run a code with $K$-fold redundancy, the global set of ranks is split
into $K$ teams. 
Each team consists of the same number of ranks, sees only the ranks of its
own team, and runs asynchronously
from all the other teams.
Each team consequently hosts one application instance of its own. 
\end{definition}

\noindent
With this definition, each rank belongs uniquely to one team.
If there are $K$ teams, each rank has $K-1$ replicas belonging to other teams.
The teams are completely autonomous, i.e.,~independent of the other teams, and
therefore consistent only in themselves:
We run \replaced{the}{each} code $K$ times and each \added{run }completes all computations and has all data.
There is neither some kind of lockstepping nor any data sharing in this
baseline version of \teaMPI.

With teaMPI, our team-based replication for SPMD is totally transparent:
An application neither does need to replicate data structures nor does it need to be aware of
the replication.
Teams are formed from subsets of MPI ranks at the simulation start-up. 
All subsequent communication calls to both point-to-point and
collective MPI routines are mapped by \teaMPI\ to communication 
within the teams.
We work with both data and computation redundancy:
Each team is a complete application instance, and,
due to SPMD, a send within team A from rank $r_1^{(A)}$ to \deleted{a }rank $r_2^{(A)}$
will 
\replaced{%
have a matching send in team B from $r_1^{(B)}$ to $r_2^{(B)}$.
}{%
at one point be matched by a send in team B from $r_1^{(B)}$ to a 
$r_2^{(B)}$.
}%

Running an application with \teaMPI\ logically means the same as using a communicator decomposition and making
each communicator run the whole simulation. 
Teams do not have to be consistent all the time. They are
\emph{weakly consistent} and fully asynchronous without added overhead for
message consistency checking.
Instead of an in-built sanity check for MPI messages, we rely on 
low-frequency consistency checks:

%
%
\begin{definition}[Heartbeat]
\replaced{%
Each rank in each team issues a heartbeat after every $\Delta t_\text{HB}$ seconds.
}{%
Each rank in each team issues heartbeats every $\Delta t_\text{HB}$ seconds.
}%
\replaced{%
Heartbeats are sent to all replica of a rank (replication multicast) 
and only carry the elapsed wall time since the last heartbeat.
}{%
Heartbeats are messages sent to all the same logical ranks of
different teams (replication multicast), and they carry only the local time
stamp.
}%
They are sent out in a fire-and-forget fashion.
 \end{definition}

\begin{figure}[btp]
 \centering%
  \includegraphics[width=0.5\textwidth]{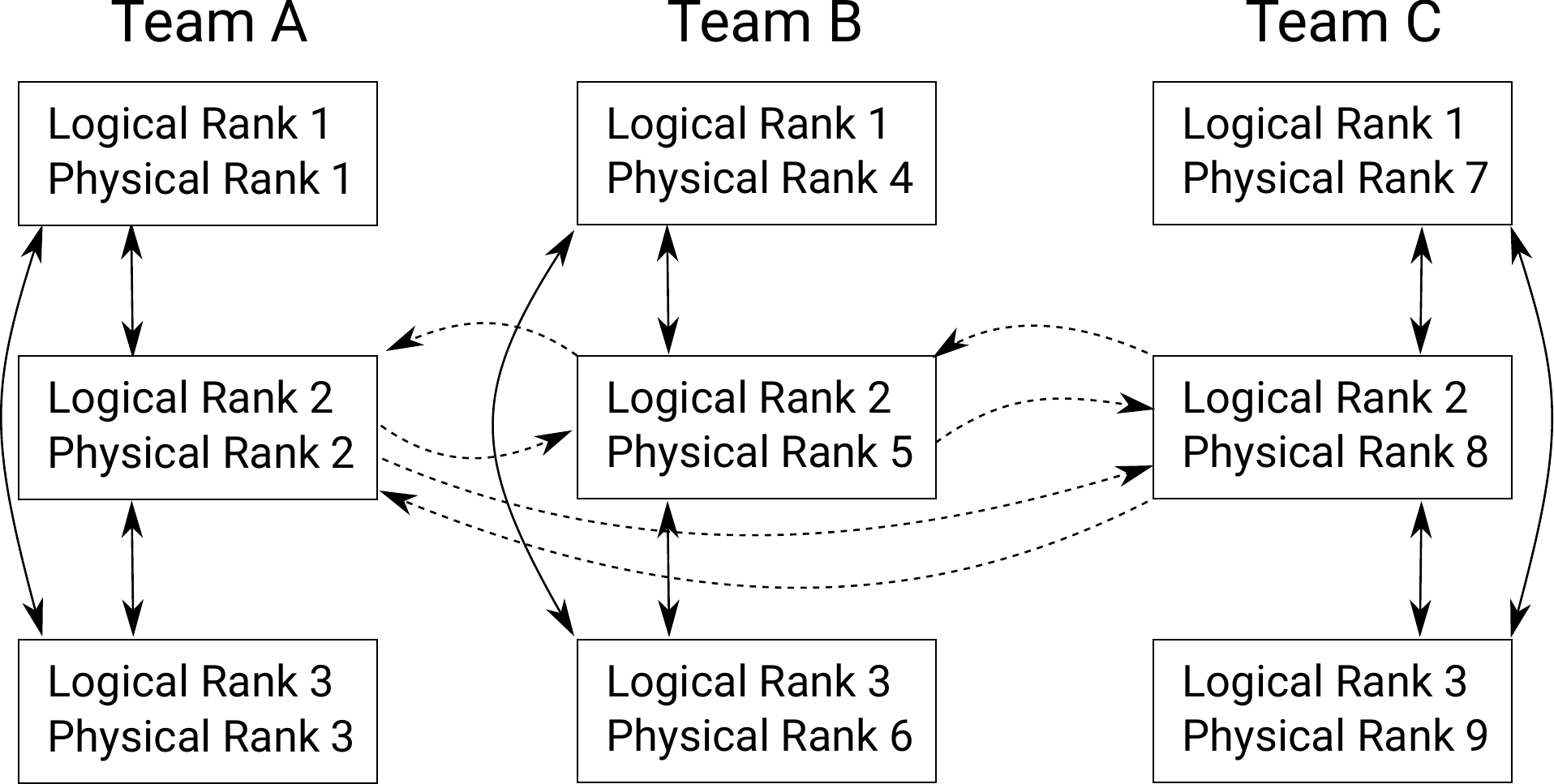}
 \caption{
  Illustration of a replication-based run with three teams, each hosting three
  ranks.
  In the baseline code, ranks communicate only with their team members (solid
  arrows), but they send heartbeats in regular
  time intervals to their replicas (dotted arrows, only illustrated for the
  logical ranks 2).
  \label{figure:teams}
 }
\end{figure}

\noindent
While all other MPI communication is restricted \deleted{so far }to intra-team only and
transfers data from the user space, heartbeats are simple non-blocking messages
exchanged between the replicas which do not carry any user data.
We clearly distinguish \emph{intra-team} from \emph{inter-replica}
communication (Fig.~\ref{figure:teams}).
The latter is hidden from the user code and is no arbitrary inter-team
data exchange.
It  only ``couples'' replicas.
As mainstream HPC machines lack support for hard real-time
scheduling, we weaken the $\Delta t_\text{HB}$ property:
We launch a task which reschedules itself all the time with a low priority and
issues 
\replaced{a heartbeat as soon as at least $\Delta t_\text{HB}$ seconds have expired since the last heartbeat task.}{heartbeats at most every $\Delta t_\text{HB}$ seconds.}

%
%
\begin{definition}[Team divergence]
A rank of a team diverges if the time in-between two heartbeats increases
compared to this in-between time in other teams.
A team diverges if at least one rank of this team diverges.
\end{definition}

\noindent
Divergence is a \emph{relative} quantity.
It results from a comparison of local
in-between times to the time stamps carried by arriving messages.
If a rank is overbooked with tasks and its heartbeat task thus is issued with   
$\Delta t \gg \Delta t_\text{HB}$, it is reasonable to assume that any replica faces
a similar delay of heartbeats as it ``suffers'' from high workload, too.
Divergence is also an \emph{observable} property: 
A rank can identify a slowing down replica, and a slow rank can identify
itself as diverging by receiving faster heartbeats from other ranks.
Finally, divergence is an \emph{asynchronous} property, as we use the timestamps written into
the heartbeats to compute in-between times. 
Local clocks do not have to be synchronised and we try to eliminate MPI message
delivery/progression effects.

Divergence nevertheless remains a \emph{statistic} quantity: As we work in
a multitasking MPI+X environment, a single late heartbeat is not a reliable
indicator that a rank is suffering from errors or overheating and thus is going
down.
If we observe divergence over a longer time span or define well-suited timeouts
$ \Delta t_\text{HB} \geq \Delta t_\text{timeout}$, we can however spot failing ranks.

Each team does exactly the same calculations and thus eventually reaches the
same state, but the heartbeats run asynchronously on-top and do not impose any
synchronisation.
They can identify that a rank is going down as it becomes slower, or
they can identify complete drop-outs.
They are ill-suited to spot data inconsistency.
However, we hypothesize that data inconsistencies eventually manifest in
corrupted data and thus in the drop-out of a complete team; which we can detect
again.

Our transparent replication
is similar to the one used in RedMPI, e.g.
\cite{fiala12detection2}. 
RedMPI and other replication models however enforce
a strong consistency model among replicated ranks,
i.e.,~make all replicas have exactly the same state subject to temporal shifts.
Individual MPI messages are double-checked against replicas
for soft errors.
This adds synchronisation. Strong consistency on the
message level furthermore becomes particularly challenging when wildcard MPI
receive operations are used.
We avoid this cliff.

To constrain overhead, many approaches do not replicate data and
computations automatically and persistently, but enable replication on-demand.
The ARRIA distributed runtime system, e.g. \cite{arria2013} schedules and
replicates tasks based on predictions about their probability of failure, while
the work in~\cite{rezaei-19-endtoend} allows to spawn resilient tasks that use
either replicated execution or checkpointing for resilience depending on the programmer's choice.
Our replication is persistent.


\section{Task sharing}
\label{section:task-sharing}

%
%
\teaMPI\ makes both data as well as all computations redundant. 
In order to save compute time, we however propose an algorithm where
teams exchange outcomes of tasks.

\begin{definition}[Shareable task]
\added{
  The tasks of interest for this paper have four important properties:
  (i) They are compute-heavy, i.e.,~exhibit a high arithmetic intensity.
  (ii) They are not a member of the critical path. We can delay their
  execution once they are spawned by some time without slowing down the
  application immediately.
  (iii) They have an outcome with a relatively small footprint relative to their
  compute cost, and the outcome is serialisable.
  We can send it around through MPI.
  (iv) They have a globally unique id.
}
\end{definition}

\deleted{For this, each task is identified by a unique id.}
\noindent
Uniqueness incorporates both the data the task is working on plus its action.
As we work with a time stepping solver, each task also is unique by the time
step it belongs to. 
Two tasks $t^{(A)}$ and $t^{(B)}$ from two different teams $A$ and $B$ thus have
the same id if and only if they perform the same action on the same data of their respective application and are
issued by the same time step.

%
%
From a user's point of view,
\teaMPI\ 
allows ranks only to exchange data within their
team, while \teaMPI\ itself exchanges heartbeats between teams.
To reduce the total cost despite the replication, we introduce
further inter-team data flow from hereon.
Both this further flow and the heartbeats however do not introduce
arbitrary point-to-point connections.
They solely remain inter-replica.

\begin{algorithm}[tbp]
  \caption{
    Wrap around the task scheduler with task sharing: We plug into the transition
    from a ready task into a running task and skip the execution if a task is already available from
    another team. In return, we send out our task results whenever we have 
    processed a task locally. This is done through a special \teaMPI\ routine,
    since \teaMPI's user interface does not support inter-team communication.
    We assume the application has equipped the task with
    MPI serialisation/deserialisation routines.
    \label{algorithm:scheduler}
  }
  \begin{algorithmic}[1]
    \Function{runTaskIfNotReceived}{task $t$}
      \State $id \gets $ \Call{computeUniqueID}{$t$}
      \If{database.contains($id$)}
       \Comment $id$ computed by other team, reuse outcome
       \State copy received outcome into task's output buffers
       \State free task outcome
       \State database.delete($id$)
      \Else 
       \Comment $id$ not yet in database
       \State \Call{runTask}{t}
       \State send $(id)$ plus task outcome to all replicas
         \label{algorithm:scheduler:send-out-task-outcome}
      \EndIf
    \EndFunction
  \end{algorithmic}
\end{algorithm}

%
%
Our extension is a
straightforward augmentation of the task runtime (Algorithm \ref{algorithm:scheduler}):
The runtime on a rank sends the outcome of any \added{shareable} task that it
has completed to all replicas, i.e.,~all the corresponding ranks in the other teams.
They receive and store them in a database.
For every ready \added{shareable} task that is to be launched, we hence validate
first that its outcome has not yet been received.
If this is not the case, we execute the actual task (and eventually distribute
its result).
If a task outcome is already in our database, we do not have to compute
the task any more.
It is sufficient to roll over the received task result and to skip the actual
computation.
To make this work, the scheduling is complemented by a receive handler
listening for task results (Algorithm~\ref{algorithm:task-drop-in}).

%
%
The database is a map from task ids onto task outcomes. 
An entry in the database indicates that a task outcome has been received\deleted{already}.
A database of received tasks as sketched so far would grow monotonically, since
tasks might drop in while they are computed.
We thus equip each database entry with a timeout and run a garbage collection regularly.
It removes all entries and cleans up buffers for tasks which are considered to
be too old.
Such a timeout could rely on heartbeat counts (``received more than x
heartbeats before'').
For explicit timestepping, it is however more convenient to use
the time step counter.
Entries older than the most recent time step won't be used anymore and can
safely be discarded.

\begin{algorithm}[tbp]
  \caption{
    Event handler that is invoked every time a task drops in from another team's
    rank.
    \label{algorithm:task-drop-in}
  }
  \begin{algorithmic}[1]
    \Function{handleTaskReceive}{task $t$}
      \State $id \gets $ \Call{computeUniqueID}{$t$}
      \If{database.contains($id$)}
       \Comment $id$ already in database, do nothing
       \State deallocate $t$
       \Comment happens if two teams compute $id$ around the same time
      \Else 
       \Comment $id$ not yet in database
       \State database.insert($id$,$t$)
      \EndIf
    \EndFunction
  \end{algorithmic}
\end{algorithm}

%
%
The algorithmic blueprint so far saves compute cost whenever a team lags
behind.
The team running ahead completes its tasks and sends out the results.
The team behind picks up the results and skips its own computations.
If two teams are roughly running in-sync, we have to modify the scheduling
slightly to benefit from the exchange between replicas:

\begin{definition}[Task shuffling]
 Let $\{t_1,t_2,t_3,\ldots\}$ be a set of tasks that are issued as ready or
 released as ready in one rush by our application.
 The first team $A$ schedules
 $\{t_1^{(A)},t_{K+1}^{(A)},t_{2K+1}^{(A)},\ldots\}$ prior to
 $\{t_{2}^{(A)},t_{K+2}^{(A)},t_{2K+2}^{(A)},\ldots\}$ and so forth.
 The second team $B$ schedules
 $\{t_{2}^{(B)},t_{K+2}^{(B)},t_{2K+2}^{(B)},\ldots\}$ prior to
 \linebreak
 $\{t_{3}^{(B)},t_{K+3}^{(B)},t_{2K+3}^{(B)},\ldots\}$ and eventually
 $\{t_1^{(B)},t_{K+1}^{(B)},t_{2K+1}^{(B)},\ldots\}$.
 This pattern continues for all teams.
\end{definition}

\noindent
Each team permutes its \added{shareable} tasks modulo the number of teams.
In practice, it is convenient to realise this through task priorities where high priority 
tasks are scheduled prior to low priority tasks.
We start from the application's task priorities but then add subpriorities with a
modulo counter which realise the shuffling.
Such shuffling even works for applications which do
not issue tasks in a batch but fire them one by one.
Shuffling weakens
the task scheduling consistency, and effectively the data consistency between
the teams.
The only situation where it might not ensure a differing task execution
ordering is when ranks issue tasks
non-deterministically.
In this case, the randomness plus the shuffling might yield similar task
execution orders for different teams.
Yet, this is unlikely.


\section{Implementation}
\label{section:implementation}

%
%
\teaMPI\ is implemented as a C++ library. 
Using the PMPI interface, \teaMPI\
intercepts the \replaced{relevant MPI calls}{MPI API calls of relevance for us} and redirects them onto
communicators or different physical ranks, respectively.
We mainly wrap blocking and nonblocking point-to-point routines as well as 
collectives.
Relying on PMPI makes 
\teaMPI\ portable.
%
%
%
%
For a replication factor $K$, an application with $R$ ranks
is started with a total of $K \cdot R$ ranks. 
This yields $K$ teams with $R$ ranks each.
Within MPI's initialisation, teaMPI 
creates subcommunicators for all intra-team communication.
Each subsequent user MPI call is hijacked by \teaMPI\ and internally mapped onto
an MPI call on the appropriate subcommunicator. 

\subsection{Implementation decisions}

\paragraph*{Heartbeats without a hard real-time environment.}

Issuing heartbeats during the simulation requires particular care.
If we make heartbeats dependent on the progression 
of the numerical simulation (for example by posting a heartbeat after every
time step), a single slow rank would delay the heartbeats of other ranks in its team:
In classical domain-decomposition approaches \replaced{(as in our example application)}{as we find it in our example
application}, point-to-point \replaced{messages }{MPI exchanges }to ``neighbour'' ranks are required before a new simulation time step can be started. 
A single slow rank 
\replaced{will therefore delay its neighbours, too.}%
{has a knock-on effect on such dependent neighbours. It will delay 
their advance, too.} 
With a heartbeat after each time step, it would thus not be possible to
isolate an individual slow or failing rank.
We could only identify \emph{teams} hosting a slow or dropped-out rank.

\teaMPI's heartbeats are issued by a special heartbeat task \added{on each rank. 
The heartbeat reschedules itself until program termination. It stores the time stamp
of the most recent heartbeat. Whenever invoked, the heartbeat task checks
whether at least $\Delta t_\text{HB}$ seconds have elapsed since the last
heartbeat. If so, a new heartbeat message is issued and the stored time stamp is
updated. We rely on two assumptions:}
\added{(i)~}Tasks run agnostic of MPI synchronization and the progression of the numerical
algorithm.
That is, even if some ranks cannot proceed with their next time step due to missing
MPI messages, they will nevertheless process the heartbeat task\deleted{ we use for
posting heartbeats around every $\Delta t_\text{HB}$ seconds}.
\added{(ii)~}\deleted{We assume that the processing time of any other task
 does not exceed $\Delta t_\text{task}$.
We can then rely on our task scheduling to call the heartbeat task frequently enough that a constant
 $\Delta t_\text{HB}$ between two heartbeats can be maintained.}
If a rank crashes\deleted{, however}, it stops issuing heartbeats.
If it slows down significantly, also its heartbeat task will be triggered 
\replaced{less often}{less
frequently}, resulting in increased time intervals\deleted{$\Delta
t_\text{task}$ } between two heartbeats.
This allows us to single out a failing rank.
\added{
 The $\Delta t_\text{HB}$ ensures that the system is not flooded with heartbeat
 messages and is not overly sensitive to small performance fluctuations
 \cite{Charrier:19:EnergyAndDeepMemory}.
}

In our implementation, \replaced{we use}{tasks build upon} Intel's
\replaced{Threading Building Blocks (TBB)}{TBB}, an abstraction from the actual
\replaced{hardware threading}{threads lacking a notion of real time}.
\added{TBB lacks support for real-time tasking.}
This introduces uncertainty. 
We do not know when exactly heartbeats are triggered.
We can not ensure that the time in-between two heartbeat sends equals
the prescribed $\Delta t_\text{HB}$.
We might even end up with situations where a rank $r^{(A)}$ sends more
heartbeats to its replica $r^{(B)}$ than the other way around.

This challenge seems to be amplified \replaced{when}{by the opportunity for} ranks \deleted{to }deploy
their results to replicas.
Any deployment moves computational load between replicas and thus,
on purpose, unbalances ranks belonging to different teams.
Real-time heartbeats would be agnostic of this.
However, their usage would contradict our
assumption that hardware failures announce themselves often through a
performance degradation.
We launch heartbeats with a fixed\added{, reasonably high} priority and rely on
the runtime to schedule the heartbeats fairly and, more importantly, roughly with the same time
intervals \added{$\Delta t_\text{HB}$} on all teams.
It is obvious that a more mature solution would use a burn-in phase without any
replication data sharing to determine a proper priority\deleted{ dynamically such
that we match roughly $\Delta t_\text{HB}$}.
The important implementation remark is that we use a comparison of local
heartbeat in-between times to the in-between times of\deleted{ the timestamps of} received
heartbeats to identify slow downs.

On the receiver side, we rely on MPI polling for the heartbeats:
Whenever a heartbeat task becomes active, we both send out our heartbeat
(multicast) and probe on available incoming ones.
If there are heartbeats in the MPI queue, we dump them into a local
array\deleted{to compare the time stamps they are carrying}.
Unexpected message arrivals should be avoided in MPI.
Yet, heartbeats do not induce a major runtime penalty.
With only the timestamp, their message footprint is small.

\paragraph*{Task Sharing.}
For the inter-team data exchange, 
\replaced{%
\teaMPI\ does not hijack standardised MPI, but offers dedicated routines.
}{%
\teaMPI\ offers dedicated routines,
i.e.~it does not hijack standardised MPI.
}%
\replaced{These}{The non-MPI} routines expose
additional inter-team communicators and \teaMPI's knowledge about the number
of teams to the application.
This way, the application can circumvent 
native MPI communication which is wrapped by \teaMPI\ and only designed for
intra-team exchange.
Task outcome sharing is multicasts sending one piece of data to all
replicas.
To use these routines, we make use of a previously developed communication infrastructure that relaxes the binding of tasks to
their spawning rank \cite{samfass2019tasks}: 
tasks and their outcomes can migrate dynamically at runtime to other processes.
For this, we add meta data (the unique ids) to the tasks and MPI
sending/receiving wrappers around both the meta data, input arguments and output.
All the wrapping is not for free:
Task outcome sharing in the resiliency context pays off if the tasks are of
reasonable workload.
We hence solely wrap compute-heavy tasks \cite{Charrier:20:EnclaveTasking}.

\paragraph*{Runtime wrapper realisation.}
Whenever a task outcome has been computed, it is buffered first before we
distribute it among the replicas. 
This is because we use non-blocking communication and the teams run
asynchronously.
A team progresses its computation irrespectively of \teaMPI's
communication. 
Without task outcome buffering, outgoing data may become inconsistent,
as the application might already overwrite the send buffer with new data.
To avoid some of the overhead of the buffering, we wrap the sends
into another condition clause:
We check before line \ref{algorithm:scheduler:send-out-task-outcome} in
Algorithm \ref{algorithm:scheduler} once more whether the task outcome has been
received already.
If so, we know that at least one replica is ahead and has multicasted
the task outcomes to the other teams while we computed locally.
There is no need to distribute the local outcome once more.

Buffering is also required on the receiver side. 
A zero-copy approach is impossible, as the receiving rank might be
slightly ahead of the sender and compute the task already itself while it receives data.
Alternatively, 
it might lag behind and might not even have created the task including
its output data fields.

\subsection{Implementation Pitfalls}
%
%
An efficient implementation of task migration between teams is technically
delicate, as we are confronted with a highly dynamic communication pattern.
\replaced{%
We cannot predict which team will be fast or how many tasks the teams exchange.
}{%
We cannot make predictions beforehand which team will be fast.
Neither can we predict how many tasks the teams exchange.
}%
Instead, we \replaced{receive }{are faced with }unexpected MPI messages, \replaced{as }{i.e.~}task outcomes
\deleted{will }arrive unexpectedly, while it  
is essential to fully overlap task sharing-related communication 
with the application and to receive shared task outcomes as quickly as possible.

\paragraph{MPI progression.}
Even though launched through non-blocking MPI, messages may actually not be
progressed internally by the \deleted{respective }MPI implementation
\cite{Hoefler:08:SacrificeThread}.
Instead, communication request handles need to be
checked for completion repeatedly through \texttt{MPI\_Test} for example. 
\replaced{%
Furthermore, standard MPI neither triggers an interrupt if messages arrive unexpectedly 
nor supports a mechanism to tell the runtime when investments into MPI progression
calls actually pay off.
}{%
Furthermore, standard MPI does not support an interrupt-based
mechanism to tell the runtime when investments into MPI progression
calls actually pay off;
neither is there an interrupt if messages arrive unexpectedly.
}%

Too little investment into MPI progression would be lethal for our task sharing approach.
We are dealing with unexpected messages which might use a rendez-vous
protocol.
If they are not detected in a timely manner, they are useless for the replicas
and even might delay the baseline application.
We therefore need \replaced{frequent}{a reasonable amount of}
\texttt{MPI\_Iprobe}s.
\replaced{Probes detect (unexpected) incoming messages and thus issue
\texttt{MPI\_Irecvs} for pending incoming tasks which consequently are
progressed through \texttt{MPI\_Test}.}
{Without, the matching \texttt{MPI\_Irecvs}
can not be started for a pending incoming task outcome.}
Shared task outcomes consequently \deleted{would not} arrive timely.

Different to our previous work where we did interweave MPI progression with 
the standard tasking \cite{samfass2019tasks},
we found it vital for \teaMPI\ to dedicate one core to an asynchronously running
communication thread \cite{Hoefler:08:SacrificeThread}, similar to our previous work~\cite{samfass18stealing,klinkenberg2020}.
It is responsible for both the progression of MPI messages (using
MPI testing),
\texttt{MPI\_Iprobe}s for detecting messages, 
as well as the progression of the task sharing algorithm (e.g.,
buffering received task outcomes and inserting them into the task outcome
database).
It ensures that task sharing actually
overlaps and is hidden from the user code.

\paragraph{Memory and communication overhead.}
Task sharing runs risk to result in an excessive memory
footprint and to yield many outstanding MPI receive and send handles.
Due to the buffering, open communication requests do not allow us to free
allocated buffers and handles.
We therefore limit the number of open send requests per process.
This effectively constrains the memory overhead and also the number of open data
exchange handles.

For explicit time stepping, this artificial limitation makes it
convenient to drop incoming tasks immediately
if they belong to a past time step.
In our algorithmic blueprint, it is rare that this happens:
If a rank is significantly ahead of a replica, it has fed the replica's
team with task outcomes which in turn makes the replica skip all task outcome sharing.
Once we limit the number of tasks, we however might run into situations where
ranks receive outdated task outcomes.
While the garbage collection would remove these as well, it is reasonable to
pipe the incoming data into a temporary buffer right away.


\section{Results}
\label{section:results}

\begin{figure}[tbp]
\centering
\includegraphics[width=0.80\textwidth]{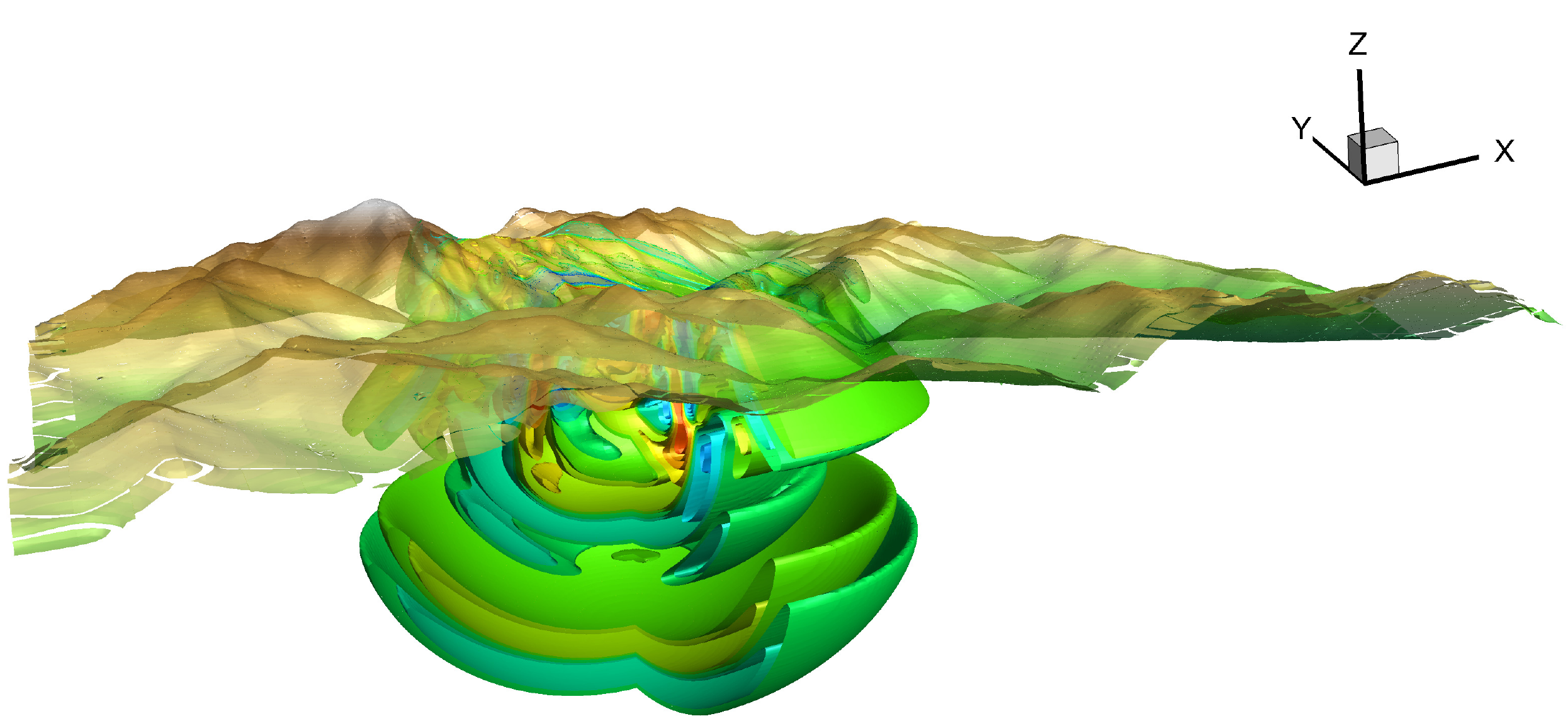}
\caption{Visualization of an example setup simulated with our code by courtesy of Maurizio Tavelli~\cite{tavelliDiffuse}.}
\label{fig:eyecatcher}
\end{figure}

Our tests are conducted on the SuperMUC-NG supercomputer operated by the Leibniz
Supercomputing Centre.
SuperMUC-NG consists of Intel Skylake Xeon Platinum 8174 nodes, where each node hosts 2$\times$24 cores\deleted{ per node}, running at a nominal clock frequency of 2.3 GHz. 
Intel Omnipath serves as interconnect.
We use \deleted{the }Intel MPI\replaced{,}{ and} Intel Compiler 2019, and Intel's TBB for
the multithreading.

We benchmark performance and functionality against a
seismology simulation for the LOH.1 setup \cite{Day:03:LOH1}.
The simulation relies on an engine for solving systems of hyperbolic
partial differential equations (PDEs) and employs an explicit, high
order Discontinuous Galerkin scheme in space and time (ADER-DG).
Its spatial discretisation stems from 
dynamically adaptive Cartesian meshes, while the code phrases its execution
in tasks relying on TBB. 
Alhough we study a benchmark, i.e.,~strip the code off many
features such as the integration of real geometries and subsurface data or
extensive postprocessing and I/O, these core features already make up a
challenging setup characterising production runs. An example visualization
obtained with our framework is shown in Figure~\ref{fig:eyecatcher}.

ADER-DG is a numerical scheme splitting up each time step into a space-time
prediction, a Riemann solve and a correction phase.
We found the prediction to be responsible for the majority of the runtime
\cite{Charrier:19:EnergyAndDeepMemory} and thus make only prediction tasks migration-ready.

One core is sacrificed to a communication thread \added{with task sharing}. 
\added{If not stated differently,} we do not take this
additional core into account when we compare the performance of \teaMPI\ with
\replaced{task sharing}{replication} to a baseline code without \replaced{task sharing}{replication}:
both the baseline as well as the task sharing versions use the same number of
cores for computations. \added{We, however, also provide data for one setup where both the baseline 
and the task sharing variant use the same number of cores per process: i.e., the task sharing 
variant uses one core less for computation than the baseline due to the core dedicated to the communication thread.}
Readers may recalibrate all \added{other} data accordingly or agree that the progression
thread is a workaround for an MPI weakness.
\subsection{Heartbeats}
%
We first demonstrate how \teaMPI\ can be used to identify failing or slow ranks with\replaced{ ExaHyPE}{ our ADER-DG PDE engine}. 
Let 56 nodes of SuperMUC-NG host two teams, each consisting
of 28 MPI ranks.  
Each rank is responsible for one part of the three-dimensional grid.
It is evenly distributed, i.e.,~the setup is load-balanced.
We configure each rank to send a heartbeat every $\Delta t_\text{HB}=1s$,
but artificially delay one rank in the first team in order to simulate a failing
node.
This is achieved by repeatedly pausing and resuming its process. 
A delay of $0.1$s kicks in after 100s.
From here, the delays increase by $0.1$s every time. 
This resembles an anticipated scenario where a failing node gradually decreases
its clock frequency before it finally goes down completely.
\teaMPI's goal has to be to identify this situation on time. 

\begin{figure}[tbp]
 \centering
 \includegraphics[width=0.55\textwidth]{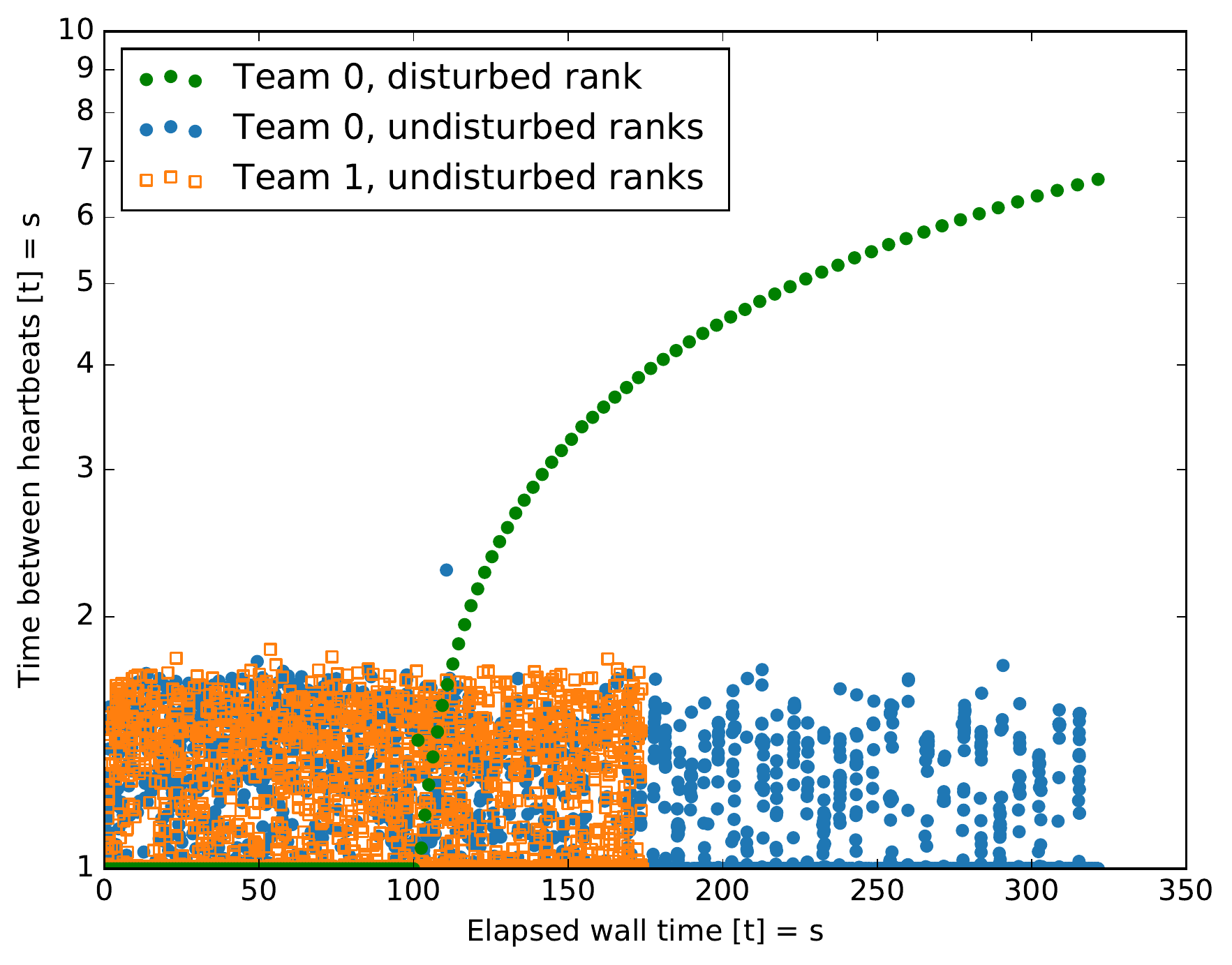}
 \vspace{-0.4cm}
 \caption{
   Time between heartbeats on 56 nodes on SuperMUC-NG if a single node is
   increasingly delayed.
   \label{fig:variable_delay}
 }
\end{figure}


%
%
A plot of the time in-between heartbeats over the first 20 timesteps (Fig.~\ref{fig:variable_delay}) unmasks the failing rank.
This rank delays its whole team 0. As a result, team 0 finishes later
and posts more heartbeats compared to the ranks in team 1.
Although one heartbeat per second is chosen, the task-based heartbeat implementation
makes the actual task timings become fuzzy and consistently exceed 1s, resulting in scheduling effects.
%
%

%
%
Every rank observes its replicas through the heartbeats. 
We cannot directly, i.e.,~in an unfiltered way, use the in-between time between
heartbeats to identify failures.
Instead, time averages have to be used to assess the healthiness of a replica.
Although a slow rank affects all members of its team (and it is thus difficult to
identify a failing rank by measuring time per time step of a team), our
heartbeats are well-suited to identify which rank is to blame for a \deleted{team}delay.


\subsection{Robustness against temporary delays}

%
%
If a single rank and, hence, team is temporarily delayed through I/O or
non-persistent hardware deteriorations (overheating) for example, 
task sharing should enable the delayed team to ``catch up'' with the faster
teams. 
To validate this hypothesis, 
we rerun the two-team setup but artificially delay the startup of one rank of
the first team:
We pause the rank for a certain time $t$ directly at startup.
To exclude stochastic effects, we make 
$t \in [45s,65s]$ uniformly distributed and
run the code with and without task outcome sharing.

\begin{figure}[tbp]
\centering
 \includegraphics[width=0.55\textwidth]{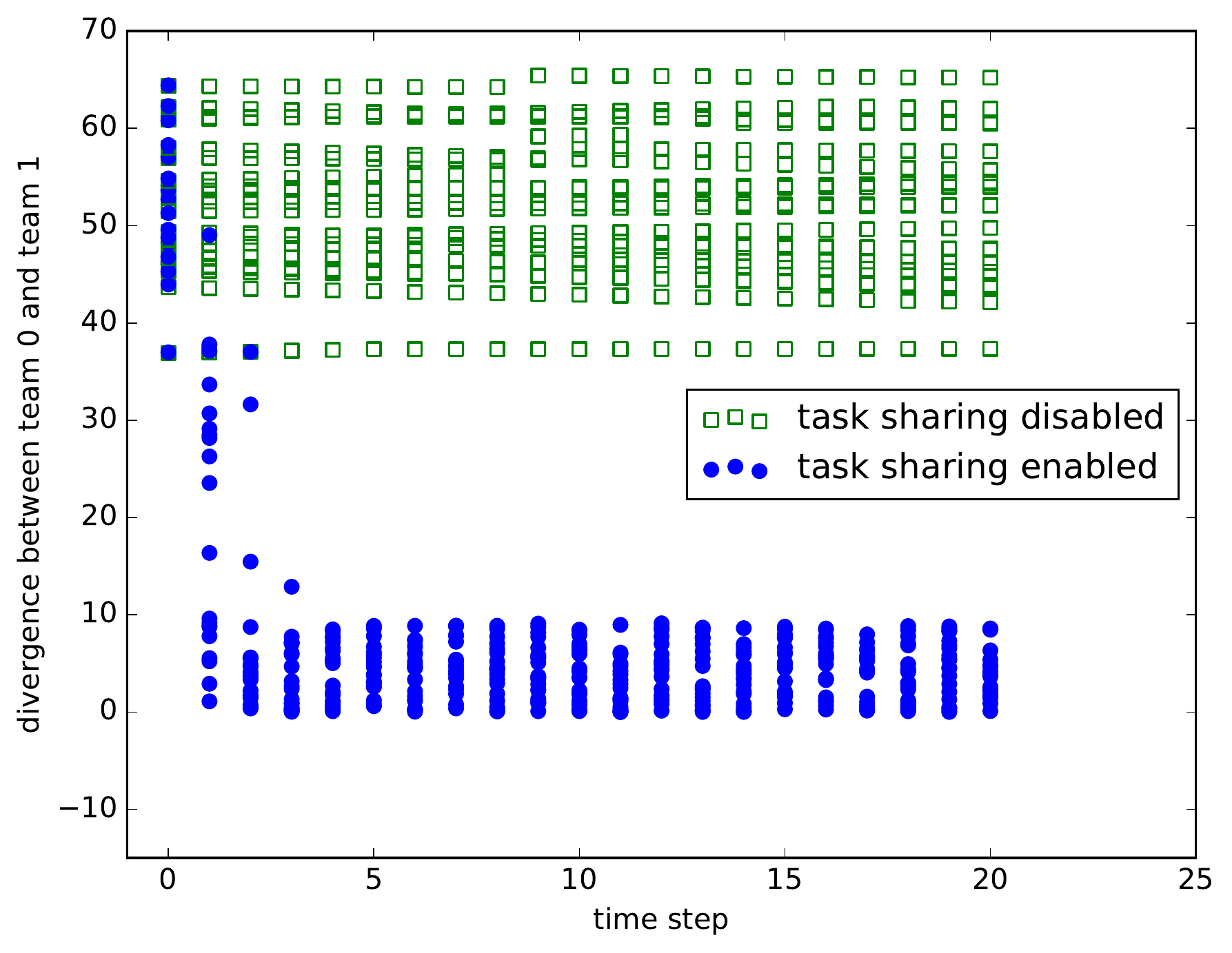}
 \vspace{-0.4cm}
 \caption{
  Team divergence for two teams for different initial delays.
  \label{fig:divergence}
 }
\end{figure}

%
%
Let $t_i^{(A)}$ be the timestamp of the start of the $i$--th timestep of team
$A$.
For teams $A$ and $B$, we can then quantify the divergence at the $i$th
timestep as $d_i^{(A,B)} = | (t_i^{(A)}-t_i^{(B)}) |$.
Without task sharing, an initial start offset between both teams persists
throughout the simulation (Fig.~\ref{fig:divergence}),
while task outcome sharing makes the divergence decrease rather quickly:
the fast team ``drags along'' the slow team, as it feeds it with task results. 

%
%

\begin{figure}[htb]
 \centering
 \includegraphics[width=0.55\textwidth]{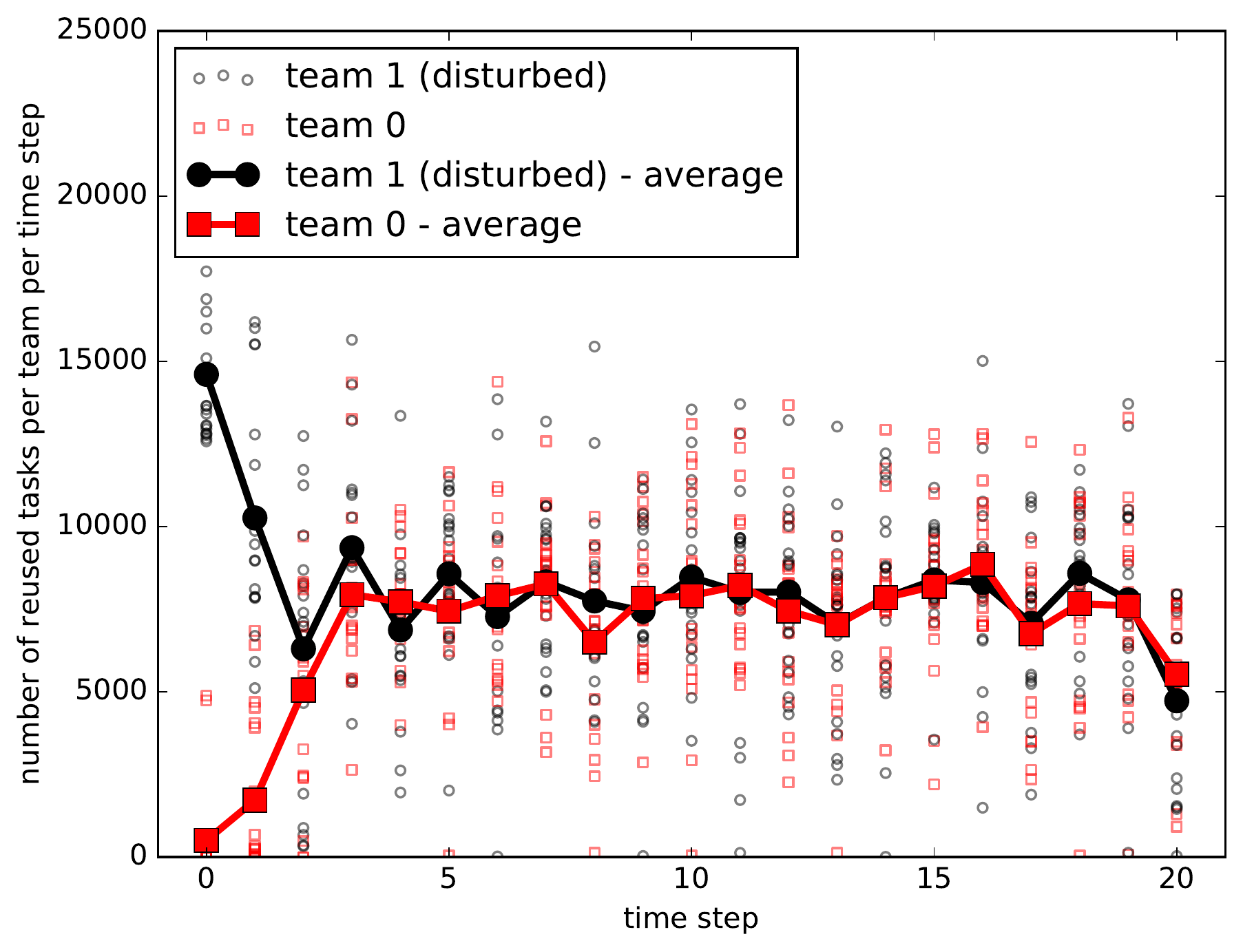}
 \vspace{-0.4cm}
 \caption{
   Number of reused tasks per team per time step for different initial random delays (with task sharing).
   \label{fig:saved_tasks}
 }
\end{figure}


We investigate this effect further 
by plotting the accumulated number of reused tasks with task sharing for
the two teams (Fig.~\ref{fig:saved_tasks}). 
Initially,
the undisturbed team reuses little to no task results from the replica team, as
the disturbed team cannot provide its results in a timely manner. 
At the same time, the disturbed team reuses tasks starting from 
the first timestep. 
It catches up.
For the delayed team, the number of reused tasks per time step \emph{decreases} over time as it catches up. Accordingly, the number of reused tasks per time step increases for the undisturbed team. Once the delayed team has catched up, the teams share tasks evenly as a result of our shuffling approach.



\subsection{Upscaling}

\begin{figure}[tbp]
\centering
\begin{subfigure}{0.49\textwidth}
  \centering
  \includegraphics[width=\textwidth]{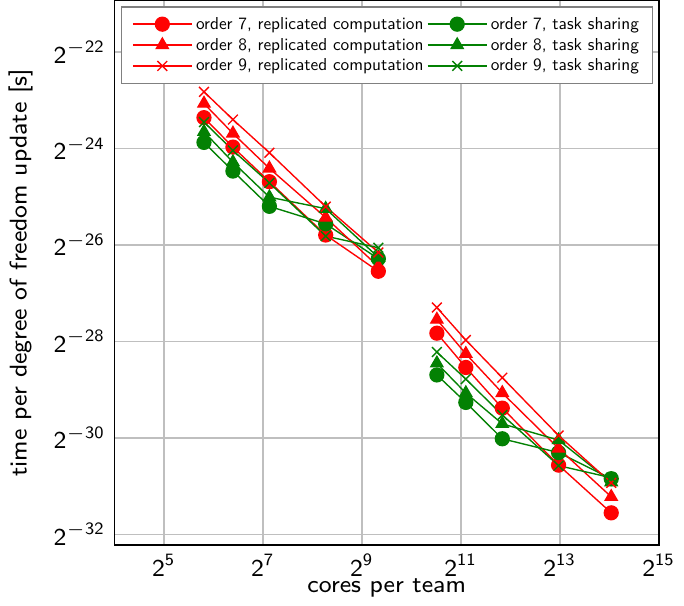}
  \caption{Same number of computation cores for baseline and task sharing.}
  \label{fig:task_sharing_normal}
\end{subfigure}%
\hspace{0.4em}%
\begin{subfigure}{0.49\textwidth}
  \centering
  \includegraphics[width=\textwidth]{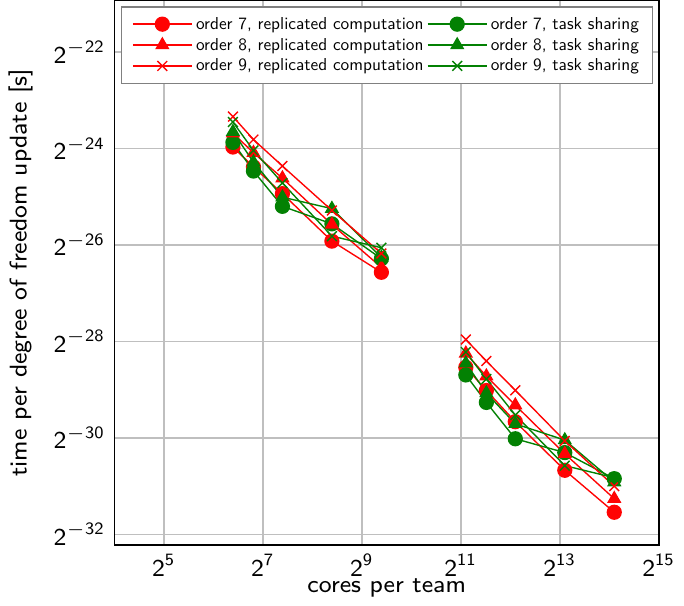}
  \caption{Baseline uses one additional core for computation compared to task sharing.}
  \label{fig:task_sharing_fair}
\end{subfigure}%
 \caption{
  Up-scaling of two teams to up to 731 nodes and $35088$ cores for varying polynomial orders: 
  Green lines show the normalized times per degree of freedom update if task
  sharing is enabled, while the red lines illustrate the vanilla
  variant where computation is done redundantly.
 }
\label{fig:task_sharing}
\end{figure}

We next study two strong scaling setups with two teams,
where we gradually increase the number of cores per rank or
team, respectively (Fig.~\ref{fig:task_sharing}). \added{We compare the task sharing 
measurements to both a baseline that uses the same number of cores for computation (Fig.~\ref{fig:task_sharing_normal})
and to a baseline that uses one additional core for computation
(the core that is sacrificed to a communication thread in the task sharing variant, Fig.~\ref{fig:task_sharing_fair})}.
We start with a domain decomposition of the computational grid that is
well-balanced, using 28 partitions for example.
These 28 partitions are mapped onto 28 ranks which are all deployed to one node.
Then we grant each rank more and more cores until the experiment eventually
spreads all 28 nodes.
We do a similar experiment with 731 ranks or partitions, respectively.
This setup eventually employs 731 nodes and $35088$ cores.
Each experiment is conducted for three polynomial orders of the underlying
Discontinuous Galerkin scheme.
The polynomial order determines how expensive the compute-heavy tasks for which
we enable task sharing are relative to the total runtime.
The higher the order the more dominant these tasks.

%
%
\replaced{Task}{\teaMPI's task} sharing yields a speedup of up to $1.5 \times$ for most
measurements. \added{In fact, task sharing can even compete with a baseline
that uses an additional core for computation in some cases, although at reduced 
speedups (compare Fig.~\ref{fig:task_sharing_normal} and Fig.~\ref{fig:task_sharing_fair}).}   
However, both experiments run into strong scaling effects at higher core counts:
If the number of cores per rank exceeds a certain threshold, the speedup induced
by \teaMPI's replication breaks down.
This breakdown occurs the earlier the smaller the polynomial order, i.e.,~the
smaller the relative cost of the shared compute tasks is relative to the total
compute time.
In the breakdown regime,
the rate of reused task outcomes decreases significantly up to the point where
hardly any computed result can be picked up by another team and all computations are
effectively replicated.
We invest twice the compute resources, but obtain the time-to-solution of a run
without any replication.

For most setups, our task outcome sharing however pays off.
Our two teams double the number of cores and thus compute cost, but we get
replication plus a significant speedup by means of walltime.
The advantageous property is lost if the balancing of cores per rank to
compute cost of the shared tasks becomes disadvantageous---which is a direct
implication of ``too many cores per rank'':
With too many, the pressure on the
communication system increases as tasks are processed and sent at a higher speed. 
The single communication thread and the interconnect can no longer sustain a fast enough transfer rate of task results. 
It just becomes cheaper to run all computations locally even though they are
done somewhere else, too.

\begin{table}[htb]
\centering
\caption{
  Total cost (in CPU hours) and speedup (in time-to-solution) with task sharing, each normalized to a single-team baseline at varying polynomial orders and number of cores per team.
  \label{table:three-teams}
}
\begin{tabular}{p{1.2cm}p{1.9cm}|rrr|rrr}
 & & \multicolumn{3}{c|}{\textbf{Total cost (CPUh)}} & \multicolumn{3}{c}{\textbf{Speedup (time-to-solution)}} \\
\textbf{Order} & \textbf{Cores/team} & \textbf{2 teams}
& \textbf{3 teams} & \textbf{4 teams} & \textbf{2 teams} & \textbf{3 teams} & \textbf{4 teams}  \\ 
\hline 
7                                    & 56           & 1.39   & 1.69   &     2.22                                 & 1.43                                          & 1.77                                          & 1.80\\
7                                    & 112          & 1.38   & 1.73   &     2.04                                 & 1.45                                          & 1.74                                          & 1.96\\
7                                    & 224          & 1.35   & 1.66   &     1.93                                 & 1.48                                          & 1.81                                          & 2.07\\
7                                    & 448          & 1.36   & 1.60   &     1.85                                 & 1.47                                          & 1.88                                          & 2.17\\
8                                    & 56           & 1.35   & 1.63   &     2.05                                 & 1.49                                          & 1.84                                          & 1.96\\
8                                    & 112          & 1.34   & 1.61   &     1.92                                 & 1.50                                          & 1.86                                          & 2.08\\
8                                    & 224          & 1.30   & 1.57   &     1.84                                & 1.54                                          & 1.92                                          & 2.18\\
8                                    & 448          & 1.30   & 1.53   &     1.71                                & 1.54                                          & 1.96                                          & 2.34\\
9                                    & 56           & 1.30   & 1.61   &     1.94                                 & 1.54                                          & 1.86                                          & 2.06\\
9                                    & 112          & 1.25   & 1.47   &     1.81                                 & 1.61                                          & 2.04                                          & 2.21\\
9                                    & 224          & 1.24   & 1.44   &     1.68                                 & 1.62                                          & 2.09                                          & 2.38\\
9                                    & 448          & 1.21   & 1.47   &     1.57                                 & 1.65                                          & 2.03                                          & 2.54
\end{tabular}
\end{table}

We continue our experimental section with further experiments where we
\replaced{use more than two teams}{compare two
team runs with three teams} (Table \ref{table:three-teams}).
The speedup behaviour persists\replaced{, yet, we }{. 
However, the walltime improvement becomes less significant with the third team
entering the game.
We} need an even higher relative compute load per task to benefit from yet
another team.
\replaced{More than three teams does not lead to any significant improvement of the time
 to solution anymore.
 As three teams are sufficient to implement resiliency where two ``valid'' ranks
 overrule the outcome of a corrupted one, we conclude that any usage of more
 than three teams is likely esoteric.
 To confirm this hypothesis, experiments with validation routines however are
 required.
 This is out of scope here.
}{ Again, higher orders hence help yet do not yield free lunch.
Overall, a replication factor beyond three seems not to be reasonable anymore.}

\begin{figure}[tbp]
\centering
\includegraphics[width=0.55\textwidth]{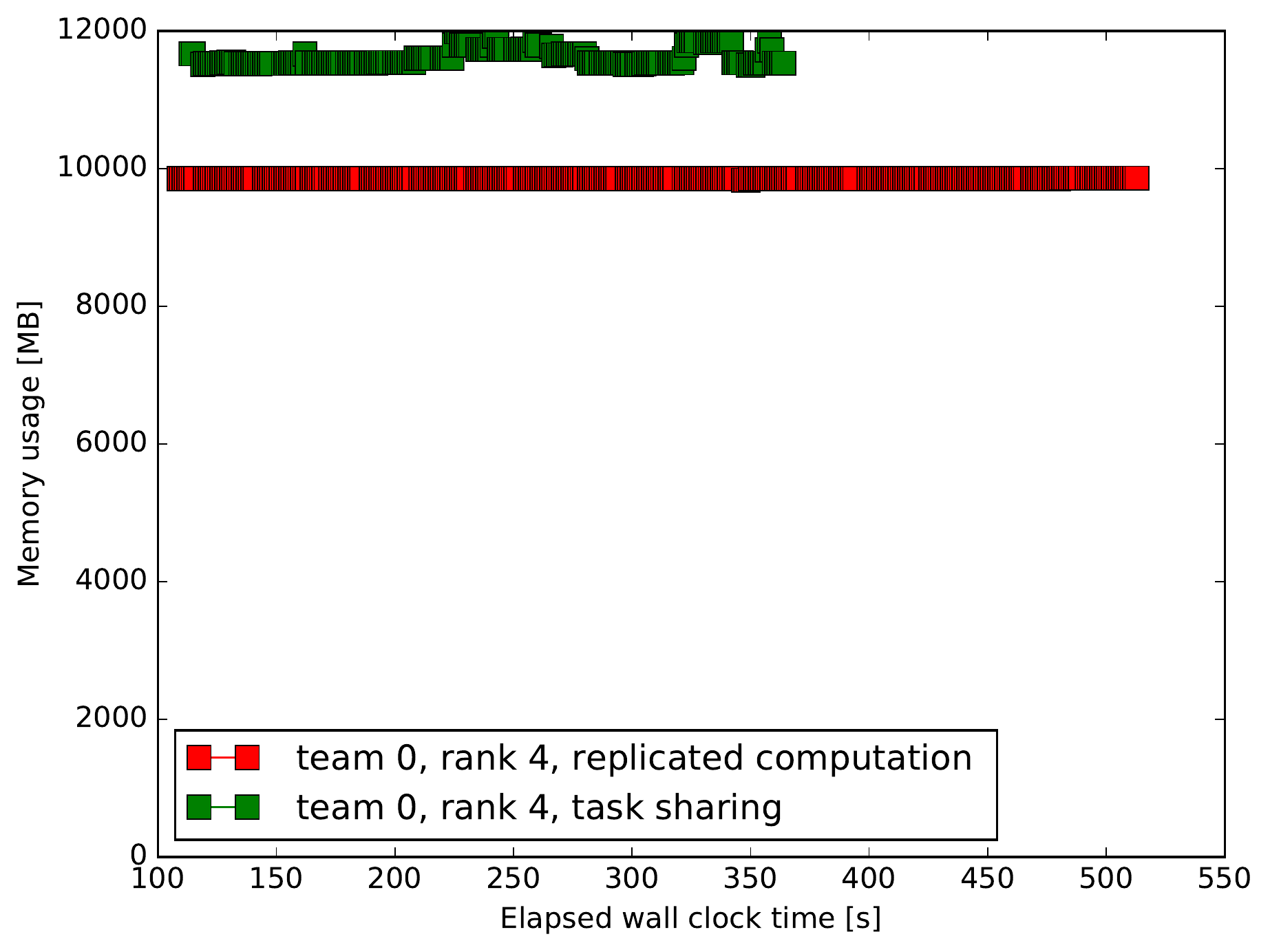}
\caption{Memory consumption of task sharing vs the baseline variant without task sharing for a selected representative rank.}
\label{fig:memory_overhead}
\end{figure}

\added{
 $k$-fold replication comes at the expense of $k$-times increased
 total memory consumption plus increased communication needs.
 On top of this, the bookkeeping of task outcomes requires further resources. 
 We quantified the
memory overhead of task sharing by repeatedly sampling each rank's memory
consumption during program execution (Fig.~\ref{fig:memory_overhead}) after the
computational grid has been allocated. In conjunction with system noise, task sharing yields a variable memory consumption pattern as task outcomes are allocated and freed dynamically. Yet, the typical additional memory overhead of task sharing
remains under control at around $20\%$ additionally used memory. 
}

\section{Conclusion and Outlook}
\label{section:conclusion}



%
%
Our paper introduces \teaMPI, an MPI wrapper/plugin which replicates a
simulation multiple times.
We call the replicas teams.
The teams run completely asynchronously.
They do however exchange heartbeats.
If the time in-between the heartbeats received vs.~the local heartbeats
diverges, we consider this to be a reliable indicator for faults.
While any rank can spot any performance degradation of a replica rank,
it is important to note that the heartbeats do not synchronise the replicas at
all and, thus, do not introduce any performance penalty.
The actual compute cost of replicas is reduced as we make each rank share its
task outcomes with the replicas which, whenever these task outcomes drop in on
time, skip their local computations and instead use the results from another
rank from another team.
This technique reduces the total CPUh cost, as long as the
computation is phrased in tasks, and as long as we do not work in the strong
scaling regime:
Enough ready tasks have to be available, so we can shuffle their order
and do not make threads idle.

%
%
Our paper introduces an elegant, minimalist and powerful new idea rendering
replication in HPC economically feasible.
It is however a conceptional piece of work.
To translate it into a production environment, we need, on the one hand, the
integration with modern MPI versions which support resiliency.
On the other hand, we have to solve \replaced{four}{three} further fundamental
challenges:
First, our code lacks a mature communication performance model for task sharing.
Specifically, it would never share too many task outcomes such that the overall
performance suffers.%
\deleted{We introduce the time-in-between heartbeats as failure alarm trigger.
However, if task outcomes are multicast, timings change.
A robust alarm mechanism should 
filter out heartbeat delays arising from task outcome migration:
\teaMPI\ should not move task outcomes around to speed up one replica and
then see the replica declare the sender of task outcomes which does all the
compute work as too late since it fails to issue heartbeats on time.}
Second, the task outcome sharing makes the whole simulation more sensitive
to soft faults (bit flips, e.g.)\added{~\cite{Reinarz:18:APosterioriFaultFrequency}}:
If a task yields an invalid outcome, this outcome might corrupt all other teams.
There is a need to develop checksums or hash techniques that can
spot such cases and veto the pollution of a run with invalid data.
The heartbeat messages might be canonical candidates to carry such
crossvalidation records.
\added{
 Third, we have to generalise our notion of shareable tasks. 
 Our strategy relies on the fact that a code yields many shareable tasks, and
 that these tasks make up a significant part of the runtime.
 To make the concept applicable to a wider range of code characteristics,
 we have to develop mechanisms that can migrate and share whole task subgraphs
 such that more fine granular tasking benefits from our
 ideas, too.
}
Finally, our approach increases the pressure on the MPI interconnects. 
It will be subject of future work to analyse how this pressure can be reduced.
To this end, we plan to investigate whether emerging technologies such as SmartNICs
can be exploited to offload the task sharing fully to the 
network hardware and to guarantee sufficient MPI progress.

\section*{Acknowledgements}

This work received funding from the European Union’s Horizon 2020
research and innovation programme under grant agreement No 671698 (ExaHyPE),
and from EPSRC's Excalibur programme under grant number EP/V00154X/1 (ExaClaw).
It used the facilities of the Hamilton HPC Service of Durham
University.
We particularly acknowledge the support of the Gauss Centre for
Supercomputing e.V.\ (www.gauss-centre.eu) for providing
computing time on the GCS Supercomputer SuperMUC at Leibniz Supercomputing
Centre (www.lrz.de).
Thanks are due to all members of the ExaHyPE consortium who made this research
possible, particularly to Dominic E.~Charrier for writing most of the engine
code and to Leonhard Rannabauer for development of the seismic models 
on top of the engine.
The underlying software, i.e.~both ExaHyPE \cite{Software:ExaHyPE} and \teaMPI,
are open source (\url{www.exahype.org} and
\url{www.peano-framework.org/index.php/teampi}).


\bibliographystyle{splncs04}
\bibliography{paper}

\end{document}